\documentclass[a4paper,10pt]{article}
\usepackage[utf8]{inputenc}

\title{AdS/CFT Correspondence Beyond its Supergravity Approximation}

\author{Mir Faizal$^1$, Ahmed Farag Ali$^{2,3}$ and Ali Nassar$^2$ \\ \\
$^1$Department of Physics and Astronomy, \\  University of Waterloo, Waterloo,
Ontario, N2L 3G1, Canada\\ 
$^2$Center for Fundamental Physics,\\ Zewail City of Science and Technology,
 12588, Giza, Egypt\\ $^3$Dept. of Physics, Faculty of Science, \\
 Benha University, Benha, 13518, Egypt. }
\date{}
\begin{document}

\maketitle

\begin{abstract}
We will study the AdS/CFT correspondence in an  intermediate region between the strong form of this 
correspondence (string theory on AdS being dual to a boundary CFT), and 
the weak form of this correspondence  (supergravity on AdS being dual to a boundary CFT). 
We  will   go beyond the supergravity approximation in the AdS by using the fact that strings have an extended structure. 
We will also calculate the CFT dual to such   string corrections in the bulk, and demonstrate that 
they are consistent with the strong form of the AdS/CFT correspondence. 
So, even though the conformal dimensions of both the relevant and the irrelevant operators   
will receive   string corrections, 
the conformal dimension of marginal operators will   not receive any such  corrections. 
\end{abstract}

\maketitle

\section{Introduction}
According to the AdS/CFT duality,  superstring theory on AdS is dual to a superconformal field theory on its boundary 
\cite{ad}-\cite{ad12}. This full correspondence is called the strong AdS/CFT correspondence. 
However, it is very difficult to analyse  this duality between the string theory 
and the boundary conformal field theory. This is because it is very difficult to analyse 
 non-perturbative aspects of the  string theory. In fact, even at 
  string tree level all aspects of string theory are not completely understood. 
So, a weaker form of the AdS/CFT correspondence is usually used. This is done 
by restricting to the analyses to the low energy effective 
field theory description of the string theory. 
It is known that the supergravity theories are low energy effective field theories  obtained from the string theory 
\cite{supera}-\cite{superb}. 
So,  according to the  weak form of the AdS/CFT correspondence,  the supergravity on AdS is   dual to 
the boundary conformal field theory \cite{super}-\cite{gsuper}.   In this paper, we will analyse  an  intermediate region
between the 
 weak form and the strong form of the AdS/CFT correspondence. We still cannot analyse the full string theory side of the 
 duality, however, in this paper, we will analyse the leading order corrections to the supergravity approximation coming 
 from the string theory.

 We will go beyond the supergravity approximation by using the fact that strings have an extended structure. 
 As strings have an extended structure, the spacetime cannot be probed below the length scale 
 of the fundamental string \cite{b}. 
 This is because the fundamental string is the smallest probe available in string theory, and 
 so it is not possible to probe anything smaller than the fundamental string. Thus, string theory naturally
 comes equipped with a minimum length scale. It is known that if  a minimum measurable length scale exists 
 in any theory, then the coordinate    representation of the momentum 
 operator will get deformed in that theory \cite{x}-\cite{2z}.  
 Furthermore, this result has been 
 generalized 
 to a covariant quantum field theory \cite{n9}-\cite{n}.
 This deformation occurs due to the  generalized uncertainty principle 
\cite{g}-\cite{g1}.  
The generalized uncertainty principle  is usually motivated from quantum gravity, 
and the Planck length is taken as the minimum measurable length scale. However, in this paper, we will use the  
 generalized uncertainty principle as a tool for 
analysing the string   corrections 
 going beyond the supergravity approximation. It may be noted that we will only be using the formalism 
 developed by generalized uncertainty principle as a tool for analysing the string   corrections, 
 and the main aim of this paper is to show that such corrections are consistent with the strong form of the 
 AdS/CFT correspondence. 
 Hence, in this paper, we will be analysing the AdS/CFT correspondence beyond its supergravity approximation. 
 
Now as the string theory comes naturally equipped with a
 minimum measurable length scale, we will expect that the low energy effective field theory derived from the string theory 
 will also be deformed because of this  minimum measurable length scale (string length scale). 
 The correction terms produced because of such a deformation will be proportional to the string length scale, 
 as they are produced because of the extended structure of strings. 
 In this paper, we will analyse such a deformation of a   field theory in the bulk. 
  We will simplify our  analysis by considering a simple scalar field theory in the bulk. 
 It may be noted that 
 even thought the field content of the full supergravity theories will be more complicated than this 
 simple scalar field theory, the basic ideas developed here can be applied to the full supergravity theory. 
 Thus, it will be possible to obtain a similar deformation for all fields in a supergravity theory. 
 As these deformations will be proportional to the length of the fundamental string, they can be 
 considered as corrections terms that will be generated by going beyond the supergravity approximation. 
 The main features of such an analysis can also be 
 understood analysing  the simple example used in this paper. 
 It may be noted that the CFT dual to a scalar field theory on the AdS has been studied in the context of  
 algebraic holography \cite{a1}-\cite{b1}. 
 
 So, in this paper, we will first deform a scalar field 
 theory by using the fact that strings have an extended structure. This deformation 
 will be proportional to 
 the length of the strings as it will occur because of the extended structure of  strings. 
 Furthermore, as these correction terms will be proportional to the length of the string, they can be considered as 
higher order correction  generated from the string theory.   
 After obtaining the corrections terms for the bulk 
 scalar field theory, we will use the AdS/CFT for calculating the boundary dual to such correction terms. 
 We will observe that 
 the conformal dimension of both the relevant and irrelevant operators will change due to such a deformation. 
 However, the conformal dimension of the marginal operators  will not receive any corrections. This is something 
 that we would expect from the AdS/CFT correspondence, according to the AdS/CFT correspondence the full string theory 
 on AdS (not only its supergravity approximation), is dual to a superconformal field theory  on the boundary.
 So, this analysis can be seen as a test of AdS/CFT correspondence going beyond the supergravity approximation.

\section{Leading Order String Corrections}
In this section, we will analyse the leading order string correction for a field theory on AdS. 
This field theory will be viewed as a part of some low energy effective field theory obtained
from string theory. 
The Polyakov action of string theory can be written as
\begin{equation}
 S = \frac{1}{4 \pi \alpha '} \int d^2 z \mathcal{L},
\end{equation}
where $\mathcal{L}$ is the Polyakov Lagrangian for   the  string theory, and the string tension $4 \pi \alpha '$ is
equal to square of the string length. The important feature of this action is that as it is the action of an extended 
body, it is not possible to use the strings described by this action to probe the spacetime  
  at distances smaller than this string length scale \cite{b}. 
 It may be noted that it is possible to obtain the conventional quantum mechanics as a low energy effective field theory 
  approximation to string theory \cite{st}. In this analysis the extended structure of strings is neglected. 
It may be noted that the extended structure of strings is not consistent with the 
usual Heisenberg uncertainty principle because  
according to the usual Heisenberg uncertainty principle,
 the length can be   measured with arbitrary precision, if the momentum is not measured.
 So, it is not possible to probe spacetime below the string length scale because of the extended structure of strings.
However, it is possible to 
 generalized the usual Heisenberg uncertainty principle to 
 a generalized uncertainty principle which   restricts the measurement of the length to a minimum value  
 \cite{g}-\cite{g1}.   This generalized uncertainty principle
deforms the usual Heisenberg algebra, and the deformation of the Heisenberg  algebra in turn deforms the 
coordinate representation of the momentum operator  \cite{5a}-\cite{54}. This deformed 
momentum operator has been   used for studying quantum field theory on spacetime with a
  minimum measurable length scale \cite{n9}-\cite{n}. It may be noted that such an analysis has been motivated from 
  quantum gravity, and hence the Planck length is taken as the minimum measurable length scale. However, in this paper, 
  we will use the formalism developed in generalized uncertainty principle, for analysing the low energy effective 
  field theory approximation to the string theory, and we will take the string
  length scale as the minimum measurable length scale. Even though the string length scale is taken to be identical with 
  the Planck length scale, there is no reason to assume that, and it is consistent to take a large length 
  scale as the string length scale. 
  It may be noted that we will only be using this formalism as a tool 
  for analysing the string theory effects going beyond the supergravity approximation. 
  So, the main motivation of this paper is to test the AdS/CFT correspondence beyond its supergravity limit. 
  
  Now we can write the  deformation of the  
  low energy Heisenberg algebra 
for a particle. Now let us start with  a   particle whose  Heisenberg algebra,
$[X, P]= i$, is deformed because of the extended 
structure of strings. 
As this deformation occur due to the extended structure of strings, so it will be proportional to 
some function of the string length which measures the  extended structure of the strings. Thus, we can write 
$
  [X, P] = i [1 + f(4 \pi \alpha') (3 P^2)], 
$
 where $f(4 \pi \alpha')$ is a suitable function of the fundamental length scale, such that 
 $f(4 \pi \alpha') \to 0$ in to limit $4 \pi \alpha' \to 0$. This is because this deformation occurs due to the extended 
 structure of strings, and so it will vanish   when we neglect such an extended structure. We will assume 
 the simplest form of such a function   $f(4 \pi \alpha') = 4 \pi \alpha'$, and write 
$  [X, P] = i [1 + 4 \pi \alpha'  (3 P^2)]. 
$ This deformation of the Heisenberg algebra causes a
deformation of the usual Heisenberg uncertainty relation 
$
 \Delta P \Delta X \geq  \left[  1 + 4 \pi \alpha' \Delta P^2 \right]/2
$.
Thus, we can write 
$
 \Delta P \leq  {(4 \pi \alpha')^{-1}}  [\Delta X \pm \sqrt{\Delta X^2 - 4 \pi \alpha'}]
$, and this 
 implies the existence of a minimum measurable length $l_{min}$ for the particle, 
\begin{equation}
 \Delta X \geq l_{min} = \sqrt{4 \pi \alpha'}.
\end{equation}
Thus, we cannot probe the spacetime below the string length scale, and so this new deformed Heisenberg algebra
is consistent with the extended structure of the strings.
It may be noted that it is possible to perform 
a similar deformation of the   usual Heisenberg algebra  in higher dimensions. So, motivated by the   
 quantum field theory on a background with an minimum measurable length scale  \cite{n9}-\cite{n}, 
the  deformation  of  the Heisenberg algebra 
  $[X^M, P_N] = i \delta^M_N$, by the string length scale can be written as 
\begin{equation}
\left[X^M, P_N \right] = i \delta^M_N  +  {4 \pi \alpha'} \left[  i \delta^M_N  g^{PQ} P_P P_Q + 2i P^M P_N\right]. 
\end{equation}
The uncertainty   relation consistent with this deformed Heisenberg algebra will again  predicts  
the existence of a minimum length   which is related to the string length, 
$l_{min} = \sqrt{4 \pi \alpha'}$ \cite{g}-\cite{n}. 
So, this deformation of the Heisenberg algebra does restrict the measurement 
of the length scale to string theory length scale. Hence, this deformation  is consistent with the existence of
an extended structure for strings. This deformation of the Heisenberg algebra also causes
 a deformation of the   momentum operators to  
\begin{equation}
 P_M^{(\alpha')} = P_M \left(1 +  {4}  \pi \alpha' g^{NP} P_N P_P\right),
\end{equation}
where, $P_N = -i \nabla_N$,
 is the original un-deformed momentum. Now   we define the operator  
 $P_M^{(\alpha')} = - i \nabla^{(\alpha')}_M$ to be the coordinate representation of the deformed momentum operator. 
 Hence, we can write 
 \begin{equation}
 \nabla^{(\alpha')}_M =  \nabla_M\left(1- {4\pi \alpha'}   g^{PQ} 
\nabla_P\nabla_Q\right). 
\end{equation}
Now let us consider a  massive free scalar field action
\begin{equation}
S=-\frac{1}{2} \int d^{d+1}x \sqrt{g}\big[g^{MN}\nabla_M\phi \nabla_N\phi + m^2 \phi^2  \big]. 
\end{equation}
We will assume this to be a part of some low effective field theory action obtained from string theory. 
Thus, we will now analyse the leading order corrections to this action that will occur due to the extended structure 
of strings. In order to do that, we will first express this action as 
\begin{equation}
S= \frac{1}{2} \int d^{d+1}x \sqrt{g}\phi \big[g^{MN}\nabla_M  \nabla_N  - m^2   \big] \phi.
\end{equation}
As this  action is thought to be a part of some low energy effective field theory derived from 
the Polyakov action of string theory, it get deformed due to the existence of the extended structure 
of strings in the Polyakov action of string theory. This deformed action can now be written as
\begin{equation}
S= \frac{1}{2} \int d^{d+1}x \sqrt{g}\phi \big[g^{MN}\nabla^{(\alpha')}_M  \nabla^{(\alpha')}_N  - m^2   \big] \phi. 
\end{equation}
Comparing the deformed and the un-deformed momentum, we can also write the equation for the deformed 
free scalar field theory as
\begin{equation}
S=\frac{1}{2} \int d^{d+1}x \sqrt{g}\phi \big[ \mathcal{D}    - m^2    
-  {8\pi \alpha'}   \mathcal{D}^2 \big] \phi,
\end{equation}
where $\mathcal{D} = g^{PQ}\nabla_P\nabla_Q$. 
  It may be noted that this is exactly the form of the action that would have been obtained 
from an derivative expansion of the massive free scalar field theory  in the framework of effective field theory 
\cite{effe}-\cite{effe1}. This is because according to the effective field theory, the low energy effective field theory 
can be written as 
\begin{equation}
S=\frac{1}{2} \int d^{d+1}x \sqrt{g}\phi \big[ \mathcal{D}    - m^2    
- \frac{1}{\Lambda}  \mathcal{D}^2 \big] \phi,
\end{equation}
 where $\Lambda$ is the scale which is integrated out. As we cannot gain any information beyond the string length scale, 
 we have to integrate it out in an effective field theory. Hence, we can identify $\Lambda^{-1} = {8\pi \alpha'} $, 
 and arrive at the same deformation using the derivative expansion in the effective field theory. 
\section{CFT Dual}
In the previous section, we obtained higher derivative corrections to the action of a scalar field theory 
due to the extended structure of strings. This was to the first order in ${4\pi \alpha'}$. 
 In this section, we will analyse the CFT dual to such 
corrections terms in the bulk. 
The equation of motion for this scalar field is given by
\begin{equation}\label{ads-gup}
(\mathcal{D} -m^2)\phi -  8\pi \alpha'\mathcal{D}^2\phi=0, \label{a}
\end{equation}
The metric for 
the  AdS   in the Poincar\'{e} patch  can be written as 
\begin{equation}
ds^2=\frac{L^2}{z^2}(dz^2+\delta_{\mu\nu}dx^\mu dx^\nu), 
\end{equation}
where  $L$ is the radius of the AdS space. 
So, the   Laplacian on AdS can be written as 
\begin{eqnarray}
\mathcal{D} &=& g^{PQ}\nabla_P\nabla_Q\nonumber \\ 
&=&  L^{-2}z^2( \partial_z^2 -(d-1)z^{-1} \partial_z+
\delta_{\mu\nu}\partial_\mu \partial_\nu). 
\end{eqnarray}
It may be noted that the term  $ 8 \pi \alpha' \mathcal{D}^2 \phi $ is generated by 
string theory effects going  beyond its supergravity approximation. This  term deforms 
the equation of motion of the scalar field  in AdS into a fourth order differential  equation.
However, the extra boundary conditions needed for it are obtained by  making  the boundary terms
appearing in the variation of the  action vanish.

Now we will obtain the solution of Eq. (\ref{a}). To do that we will use 
  the Fourier transform of $\phi (z, x)$ in $x^\mu$ coordinates
\begin{equation}
\phi(x,z)=\int\frac{d^dk}{(2\pi)^d} e^{ik\cdot x} f_k(z).
\end{equation}
We will also use   the ansatz $f_k(z)\approx z^\Delta$, for solutions close to the boundary i.e., 
we will keep only 
 the leading terms near $z=0$. It may be noted that the  conformal dimension for the boundary CFT is denoted by 
 $\Delta$, and  the boundary is defined 
at   $z = 0$.
Thus, we obtain the following equation
\begin{equation}\label{scalingdim}
\Delta (\Delta -d)-m^2L^2+\frac{8\pi \alpha' }{L^2}\big(-\Delta^4+2d\Delta^3 -d^2\Delta^2\big)=0.
\end{equation} 
It may be noted that  
we have only kept the leading order terms near $z =0$ to obtain this result. 
As the complex roots give oscillatory solutions near the boundary and this in turn makes the action un-bounded, so we will only consider real roots for this equation.
This motivates us to define  $\epsilon ( \alpha')  =8\pi \alpha' /L^2$ as  the dimensionless parameter governing the
string    correction beyond its supergravity approximation. Notice that $\epsilon ( \alpha')  \ll 1$ in the regime where we can still trust the supergravity solution in the bulk. In order for
Eq. (\ref{scalingdim}) to have only real roots, its discriminant $\Gamma$ must be bigger than zero. The  discriminant   up to order $\epsilon ( \alpha') $, is given by
\begin{equation}
\Gamma=(4 d^2 + 16 L^2 m^2) \epsilon ( \alpha')  +\mathcal{O}(\epsilon ( \alpha')  ^2).
\end{equation}
The condition $\Gamma \!>\! 0$ gives the usual Breitenlohner-Freedman (BF) bound $m^2\!>\! {-d^2}/{4L^2}$. To order $\epsilon ( \alpha')  ^2$, one obtains the following result,
\begin{eqnarray}
 \Gamma&=&-128 L^4 \epsilon ( \alpha')^2 m^4 +\big(16 L^2 \epsilon ( \alpha')
 - 32 d^2 L^2\epsilon ( \alpha')^2\big)  m^2  \nonumber  \\
&& +4d^2 \epsilon ( \alpha') + d^4 \epsilon ( \alpha')^2.
\end{eqnarray}
The condition $\Gamma>0$ can now be expressed  as
\begin{equation}
f_1(\epsilon ( \alpha')  )m^4+f_2(\epsilon ( \alpha')  )m^2+f_3(\epsilon ( \alpha')  )<0,
\end{equation}
where
\begin{eqnarray}
f_1(\epsilon ( \alpha')  )&=&128 \epsilon ( \alpha')   L^4, \\ 
f_2(\epsilon ( \alpha')  )&=&- 16 L^2 \epsilon ( \alpha') + 32 d^2 L^2\epsilon ( \alpha')^2\  \\
f_3(\epsilon ( \alpha')  )&=&-4d^2 \epsilon ( \alpha') - d^4 \epsilon ( \alpha')^2.
\end{eqnarray}
Thus, the condition $\Gamma>0$  can be written as
\begin{equation}
\frac{-f_2-\sqrt{f_2^2-4f_1f_3}}{2f_1}<m^2<\frac{-f_2+\sqrt{f_2^2-4f_1f_3}}{2f_1}.
\end{equation}
This is the modified BF bound due to the string theory effects in AdS beyond the supergravity approximation. Now, to the order $\epsilon ( \alpha')$,   we obtain the following result
\begin{equation}\label{gup-bf-bound}
-\frac{d^2(4+d^2\epsilon(\alpha'))}{16 L^2} <m^2<\frac{d^4\epsilon(\alpha')}{16L^2}+\frac{1}{8L^2\epsilon(\alpha')}.
\end{equation}
The correction to the lower bound implies that more tachyonic modes are 
allowed in AdS especially in the deep stringy regime $\epsilon(\alpha')\approx \mathcal{O}(1)$. 
The roots of Eq. (\ref{scalingdim}) can be explicitly written as
\begin{eqnarray}
\Delta_1&=&\frac{1}{2}\big(d-\sqrt{d^2+m^2L^2 }\big)+\frac{L^4 m^4 \epsilon(\alpha') }{\sqrt{d^2+4 L^2 m^2}}+\mathcal{O}(\epsilon^2), \\ 
\Delta_2&=&\frac{1}{2}\big(d+\sqrt{d^2+m^2 L^2 }\big)+\frac{L^4 m^4 \epsilon(\alpha') }{\sqrt{d^2+4 L^2 m^2}}+\mathcal{O}(\epsilon^2), \\ 
\Delta_3 &=&
\frac{1}{\sqrt{\epsilon ( \alpha') }}+\frac{d}{2}+\frac{1}{8}\left(d^2-4 L^2 m^2\right)\epsilon(\alpha')^{1/2} +
\mathcal{O}(\epsilon^{3/2}),
\\ 
\Delta_4 &=&
-\frac{1}{\sqrt{\epsilon ( \alpha') }}+\frac{d}{2}-\frac{1}{8} \big(d^2-4 L^2m^2\big) \epsilon(\alpha')^{1/2}
+\mathcal{O}(\epsilon^{3/2}).
\end{eqnarray}
Near the boundary $z=0$, a classical solution will behave as
\begin{equation}
\phi(z,x)=z^{\Delta_1}A_1(x)+z^{\Delta_2}A_2(x)+z^{\Delta_3}A_3(x)
+z^{\Delta_4}A_4(x).
\end{equation}

For the range of masses within the modified BF bound given by  Eq. (\ref{gup-bf-bound}),
we have $\Delta_1<\Delta_2$. Furthermore,   
both the roots $\Delta_3$ and $\Delta_4$ are  non-analytic in $\epsilon ( \alpha')$.
If we consider the boundary condition, $A_1(x)=A_3(x)=A_4(x)=0$,
then the boundary behavior of $\phi(z,x)$ is determined by $\Delta_2$ and 
in this case the field $\phi(z,x)$ maps to a boundary operator with conformal
dimensions $\Delta_2$.
It may be noted that  from the fact that the full string theory is dual to the superconformal field theory 
on the boundary,   the conformal 
dimension of the marginal operator dual to $\phi (z, x)$ is expected not to receive any string   corrections up to any order in $\epsilon ( \alpha')$. 
In fact, one can easily check from the explicit formula of
$\Delta_2$, that $\Delta_2(m^2=0)=d$ to all orders in $\epsilon ( \alpha')$, i.e,
the conformal dimension  does not  receive any string   
correction from effects beyond its supergravity approximation. This was expected to be
the case since $\mathcal{N}=4$ SYM theory is conjectured 
to be dual to the full string theory in the bulk according to the strong AdS/CFT correspondence, 
and the CFT operator dual to a 
massless scalar $\phi (z, x)$ in the bulk is $ {Tr}(F^2)$,
which is an exactly marginal operator in  $\mathcal{N}=4$ SYM. In this paper, we have
explicitly demonstrated that this result holds  beyond the 
supergravity approximation.
However, conformal dimensions of the irrelevant operators ($m^2>0$) and the relevant  
operators ($m^2<0$),  on the boundary CFT, do receive string 
  correction of order $\epsilon ( \alpha')$.
We can repeat this  analysis for other fields in AdS and use it for studying AdS/CFT correspondence.
We again expect that the marginal operators 
will not receive any string   corrections. 
It may be noted  that CFT operators dual to the purely stringy states 
from strong AdS/CFT correspondence are expected to scale    $ N^{2/3}$ in five dimensions, or
$  N^{1/4}$ in ten dimensions \cite{1c}. This is exactly the scaling behavior of the    new CFT 
  operators dual to the deformed  scale field theory. Hence, we can conclude that this deformation 
  does actually represent phenomena going beyond the supergravity 
  approximation. 
Finally, according to the strong AdS/CFT correspondence,  
$\mathcal{N} = 4$ SYM is dual to the full string theory on AdS, so the supersymmetry of the
boundary field theory dual cannot break by these string theory effects in the bulk theory. 
Thus, if we consider such $4 \pi \alpha'$ corrections to the   supergravity approximation of the string theory, 
we will still obtain an $\mathcal{N} = 4$ SYM on the boundary. This is because string theory effects going beyond 
supergravity approximation cannot break the supersymmetry of the boundary theory \cite{string}-\cite{string1}. So, 
the  supersymmetry cannot be broken by   the analysing done   in this paper.  We would like to point out that in this paper, 
we have only analysed the leading order string corrections, however,  even the   non-perturbative 
string corrections cannot break   supersymmetry of the boundary theory because of the strong form of the AdS/CFT correspondence. 
\section{Conclusion}
In this paper, we analysed the   AdS/CFT correspondence
beyond its supergravity approximation. We still did not analyse this correspondence for the full string theory on AdS, but 
we analysed it for the leading order corrections obtained from the string theory. 
This was done  by using that fact that strings have an 
extended structure. This extended structure of strings prevented them to probe spacetime at arbitrary 
small length scales. The existence of such an extended structure was used to deform a scalar field theory 
on AdS. Such a scalar field theory was assumed to be a part of some effective field theory approximation to 
the string theory. Even though the full supergravity action would have a very complicated field content, 
this simple example was used to understand the general features of such corrections going beyond the supergravity 
approximation. Hence, the formalism of generalized uncertainty was used as a tool to obtain string theory 
corrections to a scalar field 
theory on AdS. 
Thus, we were able to 
explicitly   obtain higher derivative corrections to the 
low energy massive free  scalar field theory on AdS. 

We also analysed the  
 boundary dual theory to this deformed scalar field theory on the bulk using the AdS/CFT correspondence. 
 As these corrections were generated from string theory, we were able to analyse the CFT dual to the string theory 
 going beyond its supergravity approximation. 
We were able to demonstrate that 
  the
conformal dimensions of both the relevant and the irrelevant operators
receive correction from these string theory effects going beyond its supergravity approximation.
However, the conformal dimension of marginal operators do not receive any such  corrections.
This was expected to occur as the 
  full string theory on AdS (not just its supergravity approximation) is dual to 
the $\mathcal{N}=4$ super-Yang-Mills  theory according to strong AdS/CFT correspondence. We also obtained the  
CFT operators are dual to  the purely stringy states. This was done by observing that they have 
the right scaling behavior.  

Finally, it will be interesting to extend this work to various  supergravity theories. 
Even though the field context of the supergravity theories would be more complicated, we expect that this procedure can 
be repeated for the full supergravity theories \cite{supera}-\cite{superb}. We also expect similar results to hold for the full supergravity action. 
However, it will be interesting to explicitly construct such actions, and demonstrate this result explicitly.
It may be noted that the UV divergences of the correlation functions of the boundary conformal field theory are related to the 
 IR divergences  of the supergravity. The
  IR divergences on the gravitational side are the same as near-boundary effects. So, they  can be analysed using the  
   holographic  renormalization \cite{4}-\cite{4a}. In this approach the 
  cancellation of the UV divergences does not depend on the IR
physics. Thus,   the holographic renormalization will only depend on
the near-boundary analysis. It would be interesting to use the holographic renormalization for analysing 
the string   corrections going beyond the supergravity approximation. 

\section*{Acknowledgments}

The work of AFA and AN is supported by the CFP at Zewail City of Science and
Technology. AFA is also supported by Benha University (www.bu.edu.eg), Egypt.
We would like to thank the anonymous referees
for  suggestions which substantially improved 
the presentation  of this  paper.


\begin{thebibliography}{99}
\bibitem{ad}  J. M. Maldacena,  Adv. Theor. Math. Phys. 2, 231  (1998).
\bibitem{da}S. Gubser, I. Klebanov, and A. Polyakov,  
Phys.  Lett.  B 428,  105 (1998). 
\bibitem{da12} A.  A.  Ardehali, J.  T. Liu and P. Szepietowski,   JHEP 6,  024 (2013). 
\bibitem{ad12} C. V. Johnson,  Nucl. Phys. B 537, 129 (1999). 
\bibitem{supera} W. Siegel,  Phys. Rev. D 47, 2504 (1993).  
\bibitem{superrrr}  K. Choi, Phys. Rev. Lett. 72, 1592 (1994). 
\bibitem{su} J. Lopez, D. Nanopoulos and A. Zichichi,  Phys. Rev. D49, 343 (1994)
\bibitem{sb} V.  Kaplunovsky and J.  Louis,  Phys. Lett. B 306, 269 (1993).  
\bibitem{superb} A. K. Tollsten,  Phys. Lett. B 300, 61 (1993).  
\bibitem{super}    L.  B. Anderson and  J. T. Wheeler,   Nucl. Phys. B 686, 285 (2004).  
\bibitem{super1} C.  Nunez, I. Y. Park, M. Schvellinge and T.  A. Tran,  JHEP 0104, 025 (2001). 
\bibitem{superg}  G. E. Arutyunov and S. A. Frolov,  Nucl. Phys. B 544, 576 (1999). 
\bibitem{gsuper} R. R. Metsaev, Phys. Lett. B 468, 65 (1999).  
\bibitem{b}D. Amati, M. Ciafaloni and G. Veneziano, Phys. Lett. B 216, 41 (1989).
  \bibitem{x}L.  N.  Chang, D.  Minic, N. Okamura, and T.  Takeuchi, Phys.Rev. D65,  125027 (2002).
\bibitem{cscds}L.  N.  Chang, D.  Minic, N. Okamura, and T.  Takeuchi, Phys. Rev. D65, 125028 (2002).
\bibitem{5a}    K. Nozari, and B. Fazlpour, Chaos, Solitons and Fractals,   34, 224 (2007).
\bibitem{za} S. Das and E. C. Vagenas, Phys. Rev. Lett. 101, 221301 (2008). 
\bibitem{n7} A. F.  Ali, S.  Das, and E.  C. Vagenas, Phys. Rev. D84, 044013 (2011).
\bibitem{14}    L. J. Garay,  Int. J. Mod. Phys. A   10,  145 (1995).
\bibitem{54}   P.  Pedram, K.  Nozari and S. H. Taheri,  JHEP 1103, 093 (2011). 
\bibitem{2z}S. Benczik, L. N.  Chang, D.  Minic, N.  Okamura, S.  Rayyan, and T.  Takeuchi,  Phys. Rev. D66, 026003 (2002).
\bibitem{n9}M.  Kober, Phys. Rev. D82, 085017 (2010).
\bibitem{skdj}V. Husain, D. Kothawala and S. S. Seahra, Phys. Rev. D 87, 025014 (2013).
\bibitem{n}M. Kober, Int. J. Mod. Phys. A 26,   4251 (2011).
\bibitem{g}   L. J. Garay, Int. J. Mod. Phys. A 10, 145  (1995).  
\bibitem{10}    P. Pedram,  Int. J. Mod. Phys. D   19,  2003 (2010).
\bibitem{5}     A. Kempf, G. Mangano, and R. B. Mann,  Phys. Rev. D   52,  1108 (1995).
\bibitem{51}    A. Kempf, J. Phys. A   30, 2093 (1997).
\bibitem{52}    F. Brau,  J. Phys. A   32, 7691 (1999).
\bibitem{g1}M. Maggiore, Phys. Lett. B 304,   65 (1993).
\bibitem{a1}K. H. Rehren, Ann. Hen.  Poin. 1, 607 (2000). 
\bibitem{b1}K. H. Rehren, Phys. Lett. B 493, 383 (2000).
 \bibitem{st}  I. Bars and D. Rychkov,  Phys. Lett. B  739 , 451 (2014). 

\bibitem{effe}  C. P. Burgess,   Ann. Rev. Nucl. Part. Sci. 57, 329 (2007).
\bibitem{2qab} T.  R. Morris, Phys. Lett. B 329,  241  (1994).
\bibitem{1qab} C.  Kim, Phys. Rev. D 74, 067702 (2006).  
\bibitem{q2ab} N.  E.  J. B. Bohr,  arXiv:hep-th/0410097 (2004). 
\bibitem{q1ab}L. Y.  Hung,   JHEP 0804, 057 (2008). 
\bibitem{effe1} T.  Banks  arXiv:1007.4001 (2010).
\bibitem{1c}O. Aharony, S. S. Gubser, J. M. Maldacena, H. Ooguri and Y. Oz,   Phys. Rept. 323, 183 (2000). 
\bibitem{string} N. Beisert, M. Bianchi, J.F. Morales and H. Samtleben,   JHEP 0402, 001 (2004).
\bibitem{stri} M. B. Green and S. Sethi, Phys. Rev. D 59, 04600 (1999). 
\bibitem{sigt} L. Anguelova, C. Quigley and  S,  Sethi,   JHEP 1010, 065 (2010). 
\bibitem{string1} K. Becker, D. Robbins and E. Witten, JHEP  1406, 051 (2014).
\bibitem{4}E. Witten,   Adv. Theor. Math. Phys. 2, 253 (1998).
\bibitem{Buchel:2009sk}
  A. Buchel, J. Escobedo, R. C. Myers, M. F. Paulos, A. Sinha and M. Smolkin,  JHEP   1003 , 111 (2010)
\bibitem{6} W. Mueck,  Nucl. Phys. B 620, 477 (2002)
\bibitem{6a}  S. de Haro, K.  Skenderis and S. N. Solodukhin,  Commun. Math. Phys. 217, 595 (2001)
\bibitem{4a} N. P. Warner,  Class. Quant. Grav. 18, 3159 (2001)
\end{thebibliography}
\end{document}